\documentclass[12pt]{iopart}

\usepackage{graphicx}
\usepackage{amsthm, amssymb}
\usepackage{setstack}

\newcommand{\binom}[2]{{#1 \choose #2}}
\newcommand{\mod}[2]{{#1 \,\,\mathrm{mod}\,\, #2}}

\begin{document}

\title[Adaptive walks on NK fitness landscapes]{Analysis of adaptive walks on NK fitness landscapes with different interaction schemes}

\author{Stefan Nowak and Joachim Krug}

\address{
Institute for Theoretical Physics, University of Cologne, Cologne, Germany
}
\ead{sn@thp.uni-koeln.de, krug@thp.uni-koeln.de}

\begin{abstract}
Fitness landscapes are genotype to fitness mappings commonly used
in evolutionary biology and computer science which are closely related to spin glass
models. In this paper, we study the NK model for fitness landscapes
where the interaction scheme between genes can be explicitly defined.
The focus is on how this scheme influences the overall shape of
the landscape. Our main tool for the analysis are adaptive walks,
an idealized dynamics by which the population moves uphill in 
fitness and terminates at a local fitness maximum. 
We use three different types of walks and investigate how their
length (the number of steps required to reach a local peak) and height
(the fitness at the endpoint of the walk) depend on the
dimensionality and structure of the landscape. We find that the
distribution of local maxima over the landscape is particularly sensitive to
the choice of interaction pattern. Most quantities that we measure
are simply correlated to the rank of the scheme, which is equal to the number
of nonzero coefficients in the expansion of the fitness landscape in
terms of Walsh functions.  
\end{abstract}


\maketitle

\section{Introduction}
In evolutionary biology, adaptation is the process by which the
genetic structure of a population changes in response
to its environment. This process
relies on two basic requirements: The supply of new individuals that
differ from the prevalent ones, and the selection of
individuals that have some kind of advantage over the others.
Differences between individuals can be ascribed to differences
in their genetic blueprint, the DNA, that are caused, e.g., by
mutation, and the advantage that is relevant for selection is an increased number of
offspring that the better adapted individuals leave in the next generation.
Instead of using four-lettered DNA sequences, 
the genotype is often represented as a binary sequence of length $L$. 
Its letters, usually taken to be 0 and 1,
are then interpreted as two different alleles that can be present
at a genetic locus.
The set of genotypes has the structure of a hypercube, a graph 
with nodes corresponding to sequences and edges connecting two
sequences when they differ by a point mutation in a single letter.
Assuming that the genotype fully specifies the reproductive success of
an individual, one may envision a mapping from the space of genotypes to the
number of offspring or some related fitness measure. 
Such a mapping is called a fitness landscape \cite{wright32,szendro13,devisser14}.

Mutations modify the genotype by changing a certain letter from zero
to one or vice versa. Whenever a new mutation arises it may become fixed, which
means that it is carried by all individuals in the population.  
The chance for this event increases with the fitness
of the new  genotype compared to the average fitness of the population
\cite{kimura62,sella05}.
If the fitness decreases due to mutation, 
fixation can only happen by stochastic fluctuations \cite{wright55}.
As the strength of these fluctuations
decreases with population size, for large
populations a mutant can survive only if its fitness is larger than average. When
additionally the rate of supply of new mutants is low such
that the timescale of fixation is much smaller than the typical time between
the appearance of different mutants, the population is monomorphic most of the time. In
this regime of strong selection and weak mutation \cite{gillespie83,orr02}
the dynamics can be approximated as an adaptive walk, in which the whole population
is treated as a single entity that travels uphill in the fitness landscape
by single mutational steps. Since the fitness has to increase in each step,
these walks terminate when there is no neighboring genotype with
larger fitness available, i.e., when a local fitness maximum has been reached.

The structure of the underlying fitness landscape is crucial for population
dynamics like adaptive walks and is often characterized
in terms of its ``ruggedness'' which can be measured, for instance, by the
number of local maxima \cite{szendro13,devisser14}.
Though there exist an increasing number of empirical fitness landscapes for small
sequence lengths~$L$ \cite{szendro13,devisser14,weinreich06,poelwijk07,lozovsky09,carneiro10,hall10,chou11,
khan11}, the mechanism by which a genotype affects
the fitness is exceedingly complex, and therefore probabilistic
models are often used for theoretical studies. In the simplest case, the fitness values are
assigned independently to each genotype from some probability
distribution, a setting referred to as the House of Cards (HoC) model \cite{kingman78,kauffman87}.

A more sophisticated model for fitness landscapes is the NK model
\cite{kauffman89,kauffman93}, which is based on the following idea:
The total fitness of a genotype is the sum of several different
contributions that are related to different properties of the individual
and depend on different parts of the genotype.
How large these parts are is controlled by the parameter $K$, while $N$ specifies the
sequence length in standard notation and hence the name of
the model (note however that we will use $L$ instead of $N$,
as the latter is often reserved for the population size).
Different parts may, and usually do, overlap and therefore one gene 
influences several contributions to the total fitness.
The pattern into which the genotype is sectioned specifies the scheme
of interaction between genes, also known as the genetic neighborhood. 

Fitness landscapes in general and the NK model in particular are also
relevant to fields outside of biology. In physics, 
the concept of an energy landscape is very similar to that of a
fitness landscape \cite{stein92}. While a population evolves into a state with
high fitness, physical systems are driven to states of low energy.
Binary sequences in particular can naturally be
interpreted as the configuration of a system with interacting spins.
In this context the HoC model is the analogue of 
Derrida's random energy model \cite{derrida81}, while the NK model
can be interpreted as a superposition of diluted $p$-spin glass models \cite{stadler99}.
In computer science, the NK model is used as a benchmark for
optimization but especially as an example for an NP-complete
problem \cite{weinberger96,altenberg97,heckendorn97,wright00}.

Among the large number of studies on the NK model
(e.g., \cite{kauffman89,perelson95,altenberg97,evans02,durrett03,limic04,waxman05,orr06,
kaul06,franke12,neidhart13}, see also section~\ref{sec:nkdef}),
only few have explicitly addressed how the choice of the interaction
scheme affects the properties of the landscapes
\cite{schmiegelt13,buzas13}. The answer to this question turns out to
depend strongly on the quantity under consideration. On the one hand,
despite earlier claims to the contrary \cite{fontana93}, the fitness
autocorrelation function is manifestly independent of the interaction
scheme \cite{campos02,campos03}. On the other hand, the accessibility of the global
fitness maximum along paths of monotonically increasing fitness
is highly sensitive to the structure of the genetic neighborhood \cite{franke12,schmiegelt13}. 

The goal of this article is to systematically  
study the influence of different genetic
interaction schemes on the landscape. 
Our main tool for the analysis
are adaptive walks, for two reasons. First, despite their simplicity adaptive walks
represent a biologically relevant limit of population
dynamics and are commonly used for the interpretation of microbial evolution
experiments \cite{schoustra09}. Second, and most importantly, adaptive walks allow for the numerical study of rather
large landscapes. Keeping in mind that a landscape consists of $2^L$ genotypes,
it is impossible to keep track of all of them when the genotype size $L$ becomes large.
Even the study of local maxima, which does not necessarily require the knowledge
of the entire landscape, becomes infeasible quickly since their relative
frequency decreases exponentially with $L$
\cite{evans02,durrett03,limic04}. Adaptive walks,
on the other hand, find local maxima rather fast, require only a tiny
fraction of the landscape to be known and are still strongly influenced
by the overall shape of the landscape, i.e., they conveniently 
translate global properties of the landscape into local ones.

The article is structured as follows: In section~\ref{sec:models} we
provide the mathematical framework and discuss the models for fitness
landscapes and adaptive walks in more detail. The results can be found
in section~\ref{sec:results} which is divided into three parts. In
section~\ref{sec:block} we study a specific interaction scheme that
enables us to derive several quantities of interest analytically,
which subsequently serve as a point of comparison to other genetic neighborhood types.
In section~\ref{sec:classic_nh} we examine numerically
the neighborhood types that are most common in the literature, and in
section~\ref{sec:clustering} we discuss the clustering of local
maxima. We then introduce the rank of an interaction scheme \cite{buzas13} as
a possible quantification of neighborhood types in
section~\ref{sec:rank} and show that most
landscape properties are correlated with it. Finally, the results are
summarized and discussed in section~\ref{sec:conclusion}.

\section{Models and methods}\label{sec:models}

\subsection{Space of genotypes and fitness landscapes}
In general, a genotype can be represented by a sequence of $L$ letters
that are drawn from an alphabet of a specific size. 
Here we will assume a binary alphabet for simplicity, i.e., each genotype
$\sigma = (\sigma_1, \ldots, \sigma_L)$ is an element of $\{0,1\}^L$.
Together with the Hamming distance
\begin{eqnarray}
 d(\sigma, \tau) = \sum_{i=1}^L (1-\delta_{\sigma_i \tau_i})
\end{eqnarray}
this can be extended to a metric space, the hypercube $\mathbb{H}_2^L$.
The distance $d(\sigma, \tau)$ is the minimal
number of mutations required to change the genotype from $\sigma$
to $\tau$ (or vice versa). A succession
\begin{eqnarray}
 \Sigma = \sigma^1 \to \sigma^2 \to \ldots \to \sigma^n
\end{eqnarray}
of genotypes is called a path, if $d(\sigma^i, \sigma^{i+1})=1$ for all $i$.

In order to quantify the reproductive value of a certain genotype $\sigma$
a fitness value $F(\sigma) \in \mathbb{R}$ is assigned to each sequence.
This mapping is called a fitness landscape.
The fitness is a measure of how well the organism is adapted to its
environment, and can be related to the (mean) number of offspring an individual
with the corresponding genotype will leave in the next generation.
A mutation from $\sigma$ to $\tau$ is called
beneficial if $F(\tau) > F(\sigma)$, and deleterious
if $F(\tau) < F(\sigma)$.
Due to natural selection, only beneficial mutations can become
prevalent in large populations.
Therefore a population undergoing adaptation propagates
through the space of genotypes along a path of monotonically increasing fitness.
Such a path, where $F(\sigma^{i+1}) > F(\sigma^{i})$ for all $i$, is called
accessible \cite{weinreich06,poelwijk07,franke11,franke12,nowak13}.

Commonly used probabilistic models for fitness landscapes are the
House-of-Cards (HoC) model, the Rough-Mount-Fuji (RMF) model \cite{franke11,aita00,neidhart14}
and the NK model. Out of these three, the HoC model is the simplest
as the fitness values $F(\sigma)$ are assigned
independently to each genotype $\sigma$ from some probability distribution.
For the HoC model, the number of maxima is particularly easy to
calculate \cite{kauffman87}: A sequence
$\sigma$ is a local maximum if and only if its fitness is larger than
that of all of its
neighbors, i.e., if it is the largest of $L+1$ random variables. The
probability for this is $1/(L+1)$ and there are $2^L$ genotypes
in the landscape, hence the expected number of maxima is
$2^L/(L+1)$.

\subsection{The NK fitness landscapes}\label{sec:nkdef}

The NK-model introduces correlations between
the fitness values of different genotypes. In this model, the
fitness $F(\sigma)$ of a sequence $\sigma$ is given by
\begin{eqnarray}
 F(\sigma)=\sum_{i=1}^L f_i(\sigma_{b_{i,1}}, \sigma_{b_{i,2}}, \ldots, \sigma_{b_{i,K}})
\end{eqnarray}
where the $f_i$ are independent HoC landscapes of size $K$, i.e., 
the $f_i(\sigma)$ are random numbers drawn independently
from the same distribution for each $i$ and each
\mbox{$\sigma \in \{0,1\}^K$}, such that a total of $L\,2^K$
random numbers are required for generating one realization of the landscape.
Unless mentioned otherwise, we will use $f_i$ 
drawn from a standard normal distribution
throughout this article, i.e., the marginal fitness of a specific genotype
is normally distributed with zero mean and variance $L$.

The $b_{i,j}$ determine the interaction between genetic loci.
For some purposes, it is more convenient to express the interaction matrix
$b_{i,j}$ in terms of neighborhood sets
\begin{eqnarray}
 V_i=\{ b_{i,1}, b_{i,2}, \ldots, b_{i,K} \}\,.
\end{eqnarray}
From a biological viewpoint there are no obvious constraints on the structure of the interaction
sets, but in the NK literature it is usually assumed that the number of sets
is equal to the sequence length $L$, that all sets contain
the same number $K$ of elements, and that $i \in V_i$ for all $i$.
The parameter $K$ is interpreted as a ruggedness parameter and interpolates
from a purely additive landscape with a single maximum ($K=1$)
to the maximally rugged HoC landscape ($K=L$).
Note that we use $K$ to denote the total number of elements in an
interaction set. This is slightly different from the usual definition, where 
$K$ is the number of elements in addition to $i$, and hence
in our notation $K$ is increased by 1 compared to the standard notation.

\begin{figure}[htb]
\centering
\includegraphics[width=0.9\textwidth]{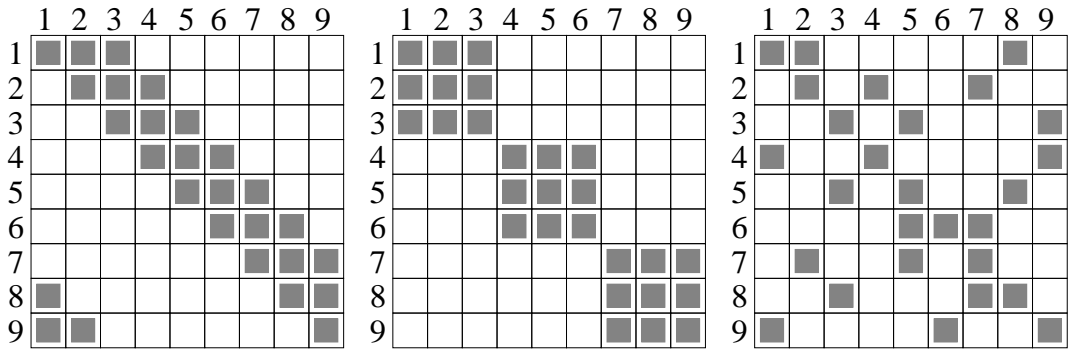}
\caption{\label{fig:nh} Illustration of the standard neighborhoods for $L=9$ and $K=3$.
From left to right: Adjacent, block, and random neighborhood.
A grey square in row $i$ and column $j$ means that $V_i$ contains $j$.}
\end{figure}

The most common types of neighborhoods,
which we are also going to use in this article, are the following (see
figure~\ref{fig:nh} for illustration). 
\begin{description}
\item[Adjacent neighborhood:] Each sub-landscape $f_i$ depends on the $i$-th locus
and its $K-1$ neighbors. The neighborhood sets are given by
\begin{eqnarray}
V_i = \{i, i+1, \ldots, i+K-1\}\,,
\label{eqn:adj_nbh_set}
\end{eqnarray}
each element modulo $L$.
\item[Random neighborhood:] The neighborhood set $V_i$ contains $i$ and $(K-1)$
other numbers, which are chosen at random from $\{1,2,\ldots,L\}$.
\item[Block neighborhood:] The neighborhood sets are given by
\begin{eqnarray}
V_i = \left\{
K \left\lfloor \frac{i-1}{K} \right\rfloor + 1, 
K \left\lfloor \frac{i-1}{K} \right\rfloor + 2, 
\ldots,
K \left\lfloor \frac{i-1}{K} \right\rfloor + K 
\right\}
\end{eqnarray}
where $\lfloor x \rfloor$ is the floor function. This means that $K$ consecutive
sets are equal, dividing the genotype into $L/K$ independent blocks ($L$ should be
an integer multiple of $K$ here). Each block
is a HoC landscape.
\end{description}

Compared to the HoC model, little is known analytically about the NK model, except for some
special cases. One example, apart from the additive case $K=1$
and the HoC case $K=L$, is the model with block neighborhood which
facilitates analytical approaches due to its modular structure \cite{perelson95,orr06}. 
As a consequence, properties like the number of maxima or the number of accessible paths to the global
maximum can be derived from results for the HoC model \cite{perelson95,schmiegelt13}.
A detailed analysis of the
NK~model with adjacent and random neighborhoods was carried out by Weinberger
\cite{weinberger91}, who derived approximate asymptotic
expressions for the number and fitness values of local maxima as well as
for the mean length of adaptive walks.
In accordance with much of the early literature on the NK-model (e.g., \cite{kauffman93}), Weinberger
concluded that these quantities are the same
for the adjacent and random neighborhoods or at least that differences between the two 
schemes are minor. 

A rigorous result for the
adjacent neighborhood model is that the mean number of
maxima grows asymptotically as $(2\,\lambda_K)^L$ with a constant 
$\lambda_K$ that increases with $K$ and depends in general on the underlying
fitness distribution of the sub-landscapes $f_i$ \cite{durrett03}.
The constant
is known exactly for a few distributions and specific, small values of $K$.
For the exponential distribution, $\lambda_2 \approx 0.5627$ and
$\lambda_3 \approx 0.6114$, for a gamma distribution with shape
parameter 2 one finds $\lambda_2 \approx 0.5646$ \cite{evans02}, and
for a negative exponential distribution
$\lambda_2 \approx 0.5770$ \cite{durrett03}. 
In \ref{sec:nummaximak1} we present a rather straightforward calculation 
to show that 
\begin{equation*}
\lambda_2 = 
 \frac{1}{6} \left[ 3 - \sqrt{3} + \sqrt{6 \, \left( \sqrt{3}-1 \right) } \right]
\approx 0.5606
\end{equation*}
for a gamma distribution with shape parameter~$1/2$.
For large $K$ and arbitrary distributions it has been conjectured that $\lambda_K$
grows asymptotically as $\exp[-\log(K)/K]$, but this was proven only
for Gaussian and fat-tailed distributions \cite{limic04}.

\subsection{Fourier decomposition of fitness landscapes}

Any fitness function $F(\sigma)$ on the hypercube $\mathbb{H}_2^L$ can be expanded into eigenfunctions
of the corresponding graph Laplacian \cite{stadler99,neidhart13,stadler96}. 
The resulting transformation is a discrete analogue of the Fourier transform 
\cite{weinberger91a} and also known as the
Walsh transform in computer science \cite{heckendorn97,weinreich13}.  
It takes on a particularly simple form
if the binary genotypes are represented by sequences
$s \in \{-1,1\}^L$ which can be interpreted as configurations of a spin system.
In this representation, the eigenfunctions of order $p$ are proportional to products
$s_{i_1} s_{i_2}....s_{i_p}$ where $0 \leq p \leq L$ and the indices $i_1, i_2,...,i_p$
are a subset of $\{1, 2,...,L\}$. For the NK-model the expansion terminates at order $p=K$,
and hence any NK fitness landscape can be written as
\begin{eqnarray}
 F(s)=F_0 + \sum_{i=1}^L H_i \, s_i + 
\sum_{p=2}^{K} \sum_{i_1 \, \ldots \, i_p} J_{i_1 \, \ldots \, i_p}
s_{i_1} \cdots s_{i_p}
\label{eqn:spin_glass_energy}
\end{eqnarray}
where the random ``magnetic fields'' $H_i$ and the ``coupling constants''
$J_{i_1 \, \ldots \, i_p}$
are determined by the original set of random functions $f_i$.
In most cases, the coefficients are extremely sparse as 
a particular coupling constant $J_{i_1 \, \ldots \, i_p}$ is nonzero only
if there is at least one neighborhood set $V$ such
that $i_j \in V$ for all $j \in \{1,\ldots,p\}$. The number of nonzero coefficients
is called the \textit{rank} of the landscape \cite{buzas13} and will be used to 
characterize different neighborhood types below in section~\ref{sec:rank}.

The Fourier spectrum of a fitness landscape is obtained from the Fourier expansion 
by summing the squared coefficients of a given order $p$. Suitably normalized, this
provides a measure for the weight of genetic interactions of different orders, and
thus a quantification of the ruggedness of the landscape 
\cite{szendro13,devisser14,weinreich13}. The Fourier spectrum
is related to the fitness autocorrelation function through a one-dimensional 
linear transformation involving discrete orthogonal polynomials \cite{stadler99}. 
For the NK-model the Fourier spectrum, like the fitness autocorrelation function, is independent of the 
genetic interaction scheme and can be explicitly calculated \cite{neidhart13}.
In contrast, as we will see in section~\ref{sec:rank}, the rank depends strongly on the choice of the genetic neighborhood.

\subsection{Adaptive walks}\label{sec:adaptive_walks}
An adaptive walk (AW) is an idealized evolutionary process. Rather than treating
the population as a set of individuals, it behaves like a single entity that
travels through the genotype space. Formally an AW is a Markov chain on $\mathbb{H}_2^L$
with dynamics defined by transition
probabilities $p(\sigma \to \tau)$ for a step from genotype $\sigma$ to $\tau$. 
Such a step is allowed only if $\tau$ can be reached from $\sigma$ by
a single beneficial mutation, i.e., if $d(\sigma, \tau)=1$ and $F(\sigma) < F(\tau)$; otherwise
$p(\sigma \to \tau)$ is zero.
This means that an AW is restricted to moving along paths that are
monotonically increasing in fitness and hence accessible in the sense
defined above \cite{weinreich06,poelwijk07,franke11,franke12,nowak13}. 
The walk terminates on some genotype $\sigma$
if no further beneficial mutations are possible, i.e., when $\sigma$ is a local
fitness maximum. The number of steps to the maximum will be
called the length $\ell$ of the AW and the fitness at the maximum will be called its 
height~$h$.

Concerning the probabilities of allowed steps, we distinguish
between different kinds of adaptive walks.
\begin{description}
 \item[Natural AW:] The transition probabilities have the fitness dependent values
\begin{eqnarray}
p(\sigma \to \tau) =
\frac{F(\tau) - F(\sigma)}{\sum_{\sigma'} [F(\sigma') - F(\sigma)]}
\,,
\end{eqnarray}
where the sum runs over all fitter neighbors $\sigma'$ of $\sigma$.
 \item[Random AW:] Each step leads to a randomly chosen fitter neighbor.
 \item[Greedy AW:] Each step leads to the fittest available neighbor.
 \item[Reluctant AW:] A step is always taken to the least fit neighbor that is still fitter than
the current genotype.
\end{description}
Note that the dynamics of greedy and reluctant walks
is completely deterministic on a given realization of the landscape.
The dynamics of the natural AW is the most realistic, in the sense that
it can be derived from individual based population models like the Wright-Fisher
or Moran model \cite{gillespie83,orr02,gillespie83a,orr98}. We will however not treat this walk type
here because its dynamics is influenced by the distribution of fitness
values \cite{neidhart11,jain11}. This is
in contrast to the other walk types, where the behavior does not depend
on the actual fitness values but only on their order. Greedy and random AWs
can be interpreted as limits of more general and
realistic dynamics \cite{seetharaman14}, 
and at least for the HoC landscape natural AW's interpolate between them
in terms of length \cite{orr02}.
The reluctant walk does not seem to have a biological
interpretation and therefore has not been considered previously in the 
biological literature, but it appears in the context of spin glasses and optimization
\cite{contucci05,contucci05.2,bussolari03,valente13}.
We will use it here as an additional tool for the analysis of fitness landscapes.

A number of results are available for random and greedy AW's on the HoC landscape.
For $L \gg 1$, the mean length of a random AW is given by
$\ell_\mathrm{HoC}^\mathrm{rnd} \approx \ln L$ \cite{kauffman87,macken89,macken91,flyvbjerg92},
while the length of the greedy AW attains a constant
limiting value $\ell_\mathrm{HoC}^\mathrm{grd} = e-1 \approx 1.7183$ \cite{orr03}.
Using the results of \cite{flyvbjerg92,orr03}, in \ref{apx:hoc_greedy} and \ref{apx:hoc_random}
we derive the mean value of the walk height for random and greedy AW's in the HoC landscape.
Assuming without loss of generality that the fitness values are uniformly distributed
on the interval $[0,1]$, the mean height is of the form 
$\langle h \rangle = 1 - \alpha / L$ to leading
order, where $\alpha$ is a constant depending on the walk type.
To our knowledge, no rigorous results are available for the reluctant 
walk, but numerically it turns out that its mean length is given by
$\ell_\mathrm{HoC}^\mathrm{rel} = L/2$ and the height constant is $\alpha=1$ (see \ref{apx:hoc_reluctant}).
A summary of the mean walk lengths and heights can be found in table~\ref{tab:hoc_walk_results}.
Since a randomly chosen maximum has an average height of
$1 - 1/(L+2)$, the fact that $\alpha=1$ for the reluctant AW implies that the
maxima found by this dynamics
are typical local maxima, whereas the random and greedy walks for which $\alpha < 1$ find
exceptionally high peaks. Moreover, on the HoC landscape the greedy AW reaches higher
fitness levels than the random AW. 
\begin{table}
\caption{Properties of adaptive walks on the House-of-Cards landscape with
uniformly distributed fitness values.
The derivation and exact values of $\alpha$ can be found in \ref{apx:height_of_walks}.
The results for the reluctant walk were obtained numerically.
\label{tab:hoc_walk_results}}
\begin{indented}
\item[]\begin{tabular}{@{}lll}
\br
Walk type & Length $\langle\ell\rangle$ & Height $\langle h \rangle = 1 - \alpha / L$\\
\mr
Greedy & $e-1$ & $\alpha=0.4003\ldots$\\
Random & $\log L$ & $\alpha=0.6243\ldots$\\
Reluctant & $L / 2$ & $\alpha=1$\\
\br
\end{tabular}
\end{indented}
\end{table}

\section{Results}\label{sec:results}

\subsection{Exact results for the block model}\label{sec:block}

In the block model, each path can be decomposed into sub-paths,
where each sub-path is confined to one specific block \cite{perelson95,schmiegelt13}.
For example, for $L=4$ and $K=2$, the path
\begin{equation*}
 \Sigma=(0011) \to (1011) \to (1001) \to (1000) \to (1100)
\end{equation*}
can be decomposed into $\Sigma^1=(00)\to(10)\to(11)$ in the first block and
$\Sigma^2=(11)\to(01)\to(00)$ in the second one.
The first mutation occurs in block 1,
the second and third mutation in block 2 and the fourth one again in block 1,
but note that any other order would also lead to a valid path with the same
endpoint.
This means that, in order to construct the full path $\Sigma$ from the
$\Sigma^i$, one also needs to know the order $\underline{\pi}(\Sigma)$
in which the blocks are affected \cite{schmiegelt13}.
However, this order has no influence on the final genotype, the length of
the path or its accessibility.

One can easily show that the probability of an adaptive step
$\sigma \to \tau$ in the full landscape,
conditioned on taking place in block $b$,
is equal to the probability of the corresponding step in the sub-landscape of
that block.
This is true under the fairly general condition that the transition probabilities
depend only on fitness differences, which applies to all adaptive walk types
defined in section~\ref{sec:adaptive_walks}.
Hence the probability that a path $\Sigma$ is taken in an adaptive walk is
given by 
\begin{eqnarray}
\mathbb{P}(\Sigma, \mathcal{L}) = \mathbb{P}(\underline{\pi}(\Sigma),
\mathcal{L}) \prod_{i=1}^{L/K} \mathbb{P}(\Sigma^i, \mathcal{L}^i)\,,
\label{eqn:aw_decomp}
\end{eqnarray}
where $\mathbb{P}(\Sigma, \mathcal{L})$ is the probability of taking
path $\Sigma$ in landscape $\mathcal{L}$, $\mathcal{L}$ is the full landscape,
$\mathcal{L}^i$ is the sub-landscape of block $i$ and
$\mathbb{P}(\underline{\pi}(\Sigma), \mathcal{L})$ is the probability
for treating the blocks in the specific order $\underline{\pi}(\Sigma)$.

As the order of blocks has no influence on the statistics we are interested in,
namely the length and height of an adaptive walk, one can treat a walk in the full landscape simply
as the succession of independent walks through the sub-landscapes. More precisely,
if $\ell_i$ denotes the random variable which represents the length of the walk
in block $i$, the length of the full walk is given by $\ell=\sum_i \ell_i$.
In the standard block model, the mean walk length is accordingly given by
\begin{eqnarray}
 \langle \ell \rangle = \sum_{i=1}^{L/K} \langle \ell_i \rangle = \frac{L}{K} \langle \ell_1 \rangle = \frac{L}{K} \ell_\mathrm{HoC}(K)
\,.
\label{eqn:nk_block_length}
\end{eqnarray}
For a random AW this leads for instance to
$\ell_\mathrm{NK}^\mathrm{rnd}\approx \frac{L}{K} \log{K}$. This
result was already obtained by Weinberger from an analysis of the density of
local maxima \cite{weinberger91}, but appeared there as an approximation for adjacent and
random neighborhoods rather than as an asymptotically exact statement for the block model.
Interestingly, according to this argument the
mean length of reluctant walks is given by $\langle\ell\rangle=L/2$
and does not even depend on~$K$. In practice, the usefulness of the relation (\ref{eqn:nk_block_length}) 
relies on an accurate knowledge of mean walk lengths on
the HoC landscape. Since the analytical expressions in table~\ref{tab:hoc_walk_results}
are only valid asymptotically for large $K$, 
we include a small-$K$ correction to $\ell_\mathrm{HoC}(K)$
that consists of two additional terms proportional to $1/K$ and $1/K^2$
with coefficients obtained by a least square fit to simulation data. 

The same argument as for the length can be used to estimate the height of an adaptive walk. We
have
\begin{eqnarray}
 h = \sum_{i=1}^{L/K} h_i\,,
\end{eqnarray}
where $h$ is the height in the full landscape and $h_i$ the height in
the $i$-th block. Since the $h_i$ have the same statistics as walk
heights in the HoC model, one can compute the mean of $h$ with the
help of previous results. 
Using additivity of the mean value as well as equation~(\ref{eqn:hoc_approx_gauss})
derived in \ref{apx:hoc_random}, we obtain 
\begin{eqnarray}
 \left< h \right> = \frac{L}{K} \left< h_1 \right>
\approx \frac{L}{K} Q^{-1}\left( 1 - \frac{\alpha \, e^{-\gamma}}{K} \right)
\,,
\label{eqn:nk_block_height}
\end{eqnarray}
where $Q$ is the cumulative distribution function of the height within a block,
$\gamma\approx0.5772$ is the Euler-Mascheroni constant 
and $\alpha$ depends on the walk type (see table~\ref{tab:hoc_walk_results}).
Note that the second approximation is only valid for fitness distributions
from the Gumbel class of extreme value theory \cite{dehaan06} and is applied here 
for a normal distribution with zero mean and variance $K$.

\subsection{Comparison of standard neighborhood types}\label{sec:classic_nh}

\begin{figure}
\centering
\includegraphics[width=0.99\textwidth]{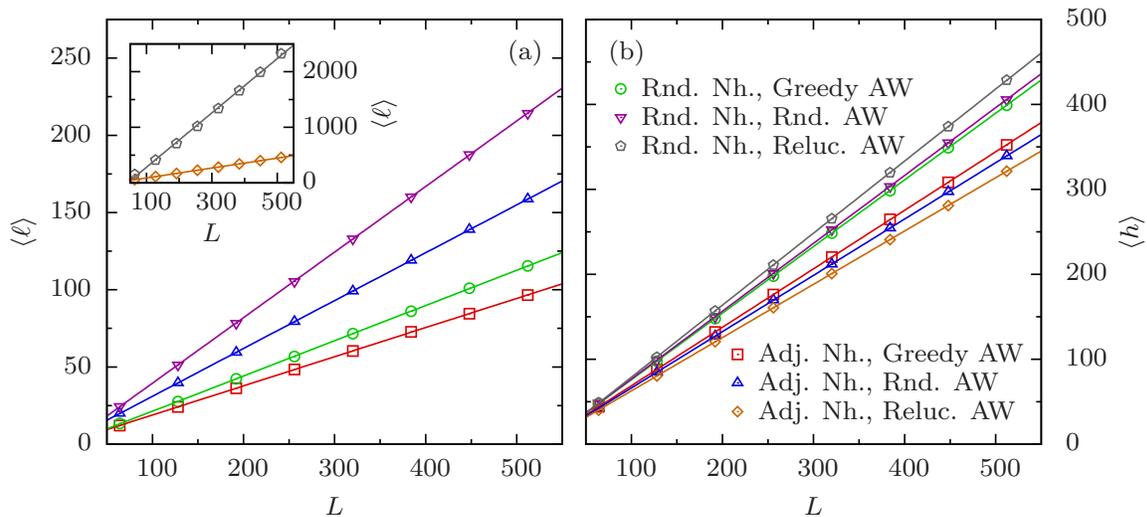}
\caption{Adaptive walk length (a) and height (b) for the NK model with 
fixed \mbox{$K=8$} and varying $L$. The symbols used 
for the different walk and neighborhood
types are explained in panel (b). Lines correspond to linear regressions. 
\label{fig:nk_varyL}}
\end{figure}

In the block neighborhood, both mean length $\langle\ell\rangle$
and height $\langle h \rangle$ of AW's are linear
in the sequence length $L$ (if $K$ is fixed),
as can be trivially seen from (\ref{eqn:nk_block_length})~and~(\ref{eqn:nk_block_height}). 
Strictly speaking, this is not true for the other neighborhood types,
since the genetic sequences cannot be divided into independent blocks
anymore. However, as shown in figure~\ref{fig:nk_varyL}, for  $L \gg K$ 
the linear behavior approximately applies for all interaction schemes.
When $L$ is comparable to $K$ the linear behavior only changes
slightly, which leads to a linear regression with almost vanishing
intercept. A notable exception is the reluctant walk on
a landscape with random neighborhoods, where the intercept is negative and
very large compared to the slope [inset of figure~\ref{fig:nk_varyL}(a)].

\begin{figure}
\centering
\includegraphics[width=0.9\textwidth]{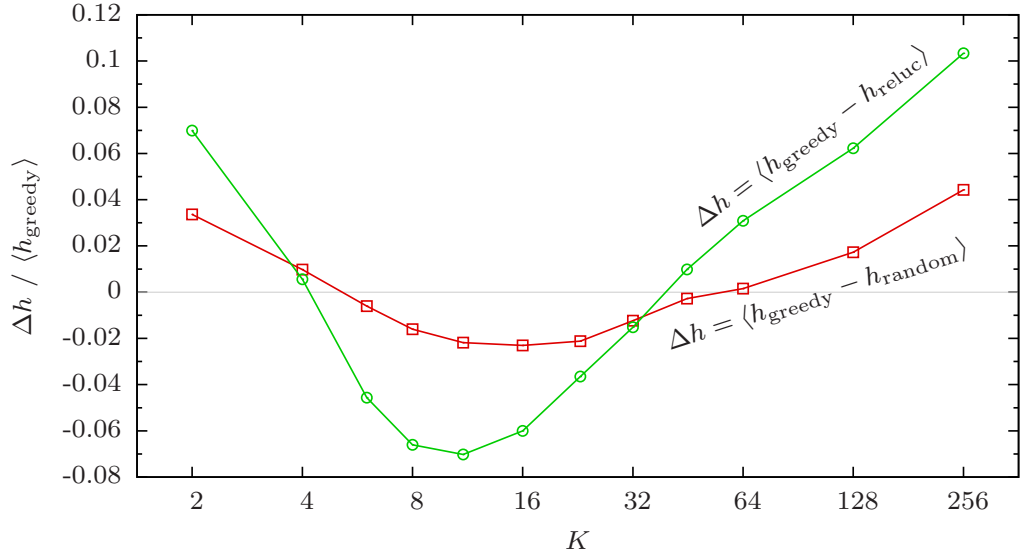}
\caption{Differences in walk height for an $L=256$ landscape with random neighborhood.
Lines are for visual guidance.
\label{fig:nk_all_grd-rnd}}
\end{figure}

The slope of the linear $L$-dependence of $\langle\ell\rangle$ and $\langle h \rangle$
in figure~\ref{fig:nk_varyL} differs markedly between
different neighborhood and walk types. As in the HoC model, the greedy
walk has the shortest length of all the walk types, the reluctant walk
is the longest and random adaptive walks are in between.
The neighborhood type has an influence on the length which is comparable in
strength to that of the walk type. For a given walk type, random neighborhoods facilitate longer
walks than adjacent neighborhoods, which in turn give rise to longer walks
than the block neighborhood. The influence of the neighborhood on walk length
is most pronounced for the reluctant walk [see the inset of
figure~\ref{fig:nk_varyL}(a)]. 
The ordering of neighborhood types remains the
same if one looks at the walk height instead of length, 
but the situation regarding the different walk types is more complex
[figure~\ref{fig:nk_varyL}(b)]. While for the 
adjacent and block neighborhood the height increases with the ``greed''
of the walk, the order is reversed for the random neighborhood.
However, this is not the case in general but only for
suitable choices of $K$ (see also figure~\ref{fig:nk_all_grd-rnd}). 

\begin{figure}
\centering
\includegraphics[width=0.99\textwidth]{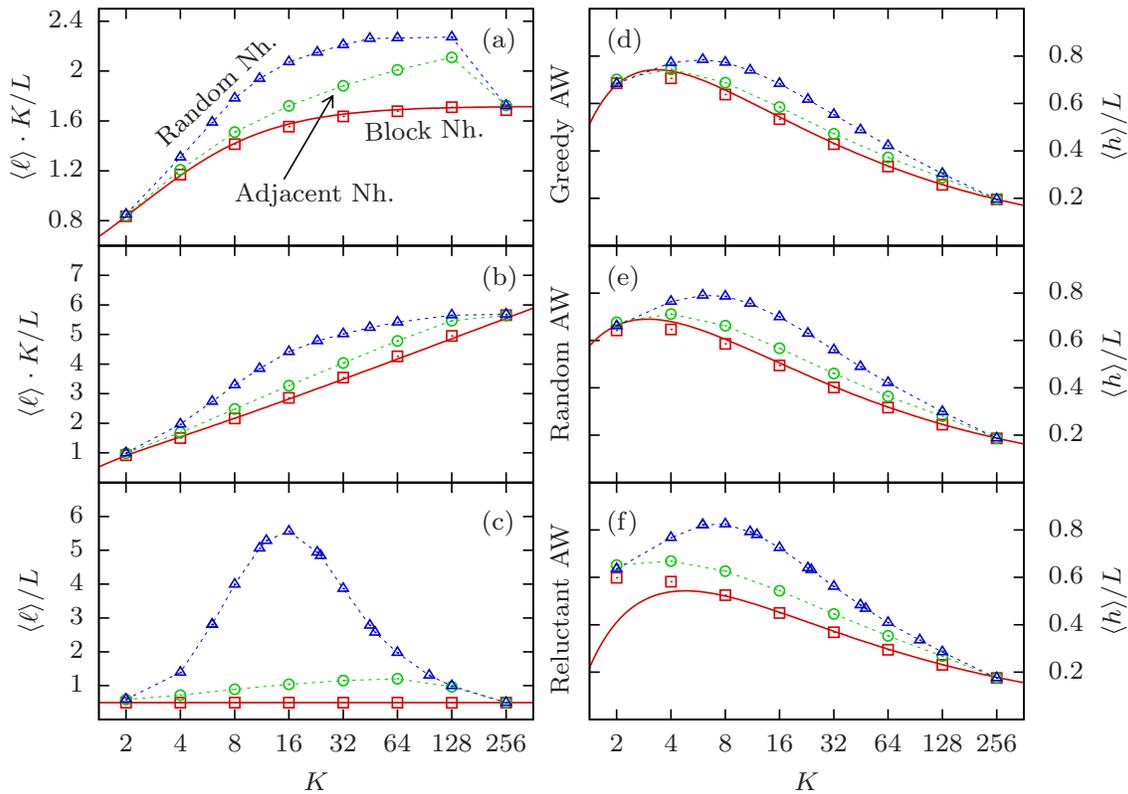}
\caption{Adaptive walk length (a)-(c) and height (d)-(f) 
for the NK model with fixed \mbox{$L=256$} and varying $K$.
Note that different scalings of the mean lengths $\langle \ell \rangle$
have been applied in panels (a)-(c) 
in order to emphasize the differences between neighborhood types.
Solid lines correspond to the analytical expressions for the block neighborhood given by~(\ref{eqn:nk_block_length})
and~(\ref{eqn:nk_block_height}), respectively. Symbols are explained in panel (a), 
and dashed lines are for visual guidance.
\label{fig:nk_all}}
\end{figure}

Both lengths and heights of AW's 
depend sensitively on $K$, with the length of reluctant walks in the block
model being the only exception.
Figure~\ref{fig:nk_all} shows the dependence on $K$ for different choices
of the neighborhood and walk type.
For all values of $K$ (except $K=1$ and $K=L$)
both walk length and height are consistently largest with
random neighborhood, second largest with adjacent neighborhood
and smallest with block neighborhood.
The difference between 
neighborhood types is most apparent for intermediate values of $K$. This
is not surprising, because for $K=1$ and $K=L$ all neighborhood types are
equivalent and the behavior is expected to change smoothly with $K$.

Since the local maxima of the fitness landscape are the absorbing states for
adaptive walks, their number $N_\mathrm{max}$ should be inversely correlated
with the length of adaptive walks. In agreement with other
studies \cite{schmiegelt13,buzas13}, our findings thus suggest that, 
for given values of $L$ and $K$,
$N_\mathrm{max}$ is largest in a landscape with block neighborhood,
slightly decreased for adjacent neighborhood and the smallest
for random neighborhoods.
Moreover, assuming that the number of maxima generally increases with $K$,
this should result in a decrease of $\langle\ell\rangle$ which can indeed be observed
for greedy and random AWs (but note that this is not visible in figure~\ref{fig:nk_all}
because of the scaling of $\langle\ell\rangle$).
However, reluctant walks show an unexpected departure from this pattern.
The reluctant walk length
is constant in $K$ in the block model and displays a non-monotonic behavior
for adjacent and random neighborhoods [figure~\ref{fig:nk_all}(c)]. 
In particular, the combination
of the reluctant walk and random neighborhoods results in extremely long
walks with a length that is several times larger than the diameter $L$
of the genotype space. This implies that on average each site in the sequence mutates
several times before a local maximum is reached. 

Similarly, 
the height of adaptive walks should be related to the height of local
maxima. In previous work it was found that the height of an
average local maximum in the NK-model decreases asymptotically as
$\sqrt{\log(K) / K}$  for large $K$ \cite{durrett03,weinberger91}.
The relevance of this effect in the present context should however not be 
overestimated, since the results described in section~\ref{sec:adaptive_walks} for the 
HoC landscape show that adaptive walks generally do not terminate on random
maxima but on particularly high ones. As more maxima become available
with increasing $K$, the walks might find higher ones even though
their average height decreases. Be that as it may, the resulting dependence
of $\langle h \rangle$ on $K$ is not monotonic and has a maximum at
rather small values of $K$ [figures~\ref{fig:nk_all}(d)-(f)]. By changing the neighborhood from
block over adjacent to random, this maximum becomes more pronounced
and is shifted slightly to larger $K$.

Concerning the different walk types, the behavior of $\langle h \rangle$
looks qualitatively similar at first glance.
However, when comparing the walk types on the same
landscape model a more interesting picture emerges. 
In the block model greedy walks attain a larger
height than random AW's and reluctant walks reach the lowest heights, as would
be expected from the results on the HoC landscape. For adjacent
neighborhoods, the order of the heights remains the same, but the differences
become smaller.
However, for random neighborhoods, one can find values of $K$ where
this order is reversed, implying that the reluctant walks are most efficient in 
locating high fitness values (see figure~\ref{fig:nk_all_grd-rnd}).
This effect was also observed previously for similar landscape models
in a different context \cite{bussolari03,valente13}.

\subsection{Clustering of local maxima}\label{sec:clustering}

In addition to the number of local fitness maxima, also their distribution in sequence
space should be expected to affect the behavior of adaptive walks. 
Even on the uncorrelated HoC landscape, the probability that
a randomly chosen genotype is a local maximum
is given by $p_\mathrm{max}=1/(L+1)$ while the probability that two genotypes
at distance 2 are both maxima is given by
$p_\mathrm{max,2}=1/[L\,(L+1)] > p_\mathrm{max}^2$ \cite{Baldi89},
i.e., in the proximity of a local maximum it is more probable to find another.
This effect is weak on the HoC landscape,
but on a correlated landscape the clustering of maxima can become quite pronounced, as
has been repeatedly noted in the NK literature \cite{kauffman93,ostman14}.

\begin{figure}
\centering
\includegraphics[width=\textwidth]{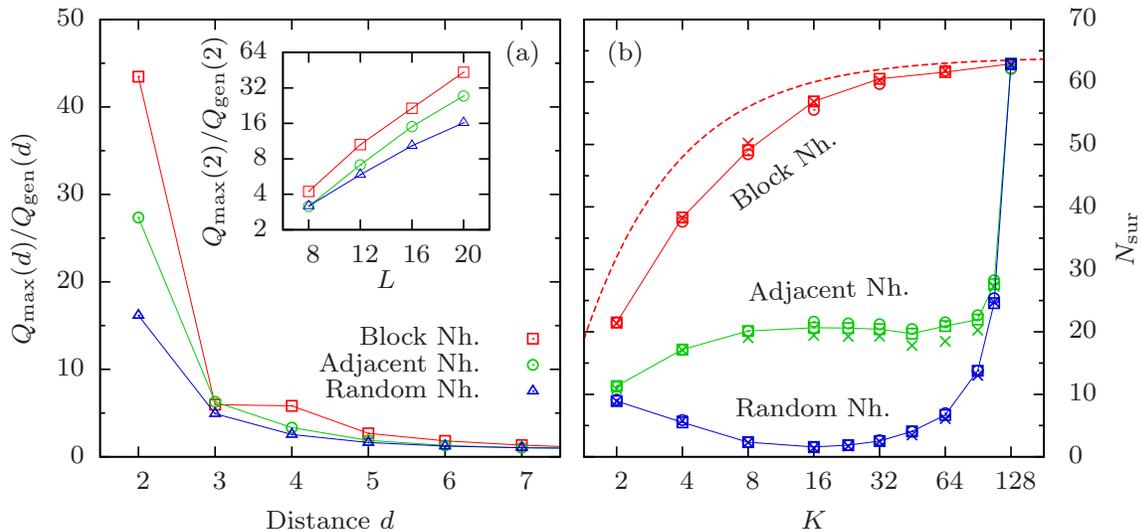}
\caption{Clustering of local maxima.
(a) Distribution $Q_\mathrm{max}$ of the distance $d$ between pairs of local
maxima, normalized to the corresponding distribution $Q_\mathrm{gen}$ of all genotypes.
Landscape parameters are $L=20$ and $K=5$. Inset shows
$Q_\mathrm{max}/Q_\mathrm{gen}$ at distance $d=2$ as a function of $L$
for fixed $L/K = 4$.    
(b) Mean number $N_\mathrm{sur}$ of local maxima that are at distance $d=2$ to the final
genotype of an adaptive walk on an $L=128$ landscape.
Circles correspond to greedy, squares to random and crosses to reluctant walks,
though the walk type does not
have a large impact on the result. Solid lines are for visual guidance.
The dashed line shows $N_\mathrm{sur}$
according to~(\ref{eqn:maxima_clustering}) for a randomly chosen
maximum in the block model. In the limiting case $K=L$ corresponding
to the HoC model, $N_\mathrm{sur} = (L-1)/2$ independent of the
interaction scheme. 
\label{fig:nk_clustering}}
\end{figure}

For the block model, this effect is easy to quantify.
A genotype $\sigma$ is a fitness maximum, if and only if its blocks
correspond to maxima in their sub-landscape.
Hence the corresponding probability $p_\mathrm{max}$ is given by
\begin{equation*}
 p_\mathrm{max} = \left( \frac{1}{K+1} \right)^\frac{L}{K}
\,.
\end{equation*}
Now let $\tau$ be a second genotype that is randomly chosen
under the constraint $d(\sigma, \tau)=2$. In order to be a
local maximum as well, the loci in which $\tau$ differs from
$\sigma$ have to be within the same block, which is fulfilled
with probability $(K-1)/(L-1)$, and there must be 
another local maximum at this position in the block, which is true
with probability $1/K$. Therefore, the probability $p_\mathrm{max,2}$ 
is given by
\begin{equation}
 p_\mathrm{max,2} = p_\mathrm{max} \, \frac{K-1}{K\,(L-1)}
\label{eqn:maxima_clustering}
\end{equation}
which is vastly larger than $p_\mathrm{max}^2$ for sufficiently large $L$.
The clustering of local maxima can also be observed for the other neighborhood
types. In figure~\ref{fig:nk_clustering}(a) we display the distribution
of distances between local maxima, showing that the clustering of maxima 
is strongest for the block
neighborhood while it is weakest for the random neighborhood.

The analysis in figure~\ref{fig:nk_clustering}(a) was 
restricted to rather small landscapes of size $L=20$ where it is feasible
to exhaustively sample all genotypes. 
For much larger landscapes this is no longer possible and 
it is very difficult to devise an unbiased search algorithm that
randomly samples local maxima. 
For this reason, we simply consider
local maxima that were found by an adaptive walk and
determine the mean number $N_\mathrm{sur}$ of maxima surrounding such a maximum, i.e., those at 
the minimal distance $d=2$.
The result is shown in figure~\ref{fig:nk_clustering}(b).
Apparently, the walk type does not have a large impact on $N_\mathrm{sur}$,
but the neighborhood type does. For intermediate values of $K$, the number
of surrounding maxima in the random neighborhood
differs from the results for block neighborhoods by a factor of almost 50,
while the results for the adjacent neighborhood lie, as always, roughly halfway
between block and random neighborhood. To assess how strongly these results
are biased by the sampling of the maxima by an AW, one may compare the numerical
results for the block model to the corresponding prediction for randomly 
chosen local maxima derived from~(\ref{eqn:maxima_clustering}). It is seen
in figure~\ref{fig:nk_clustering}(b) that $N_\mathrm{sur}$ is slightly larger for 
randomly chosen maxima, implying that the maxima found during
an AW are more isolated than the typical ones. Nevertheless, the effect of 
sampling bias appears to be rather minor, and we conclude that
the study of $N_\mathrm{sur}$ exposes one of the most
recognizable differences between neighborhood types.

\subsection{Neighborhood rank}\label{sec:rank}

The rank $r$ of NK-neighborhoods was introduced in \cite{buzas13} to quantify
neighborhood schemes, and it was shown numerically to be
negatively correlated to the number of maxima of a landscape if $K$ and $L$
are kept constant. The rank of an NK neighborhood scheme is defined as
\begin{eqnarray}
 r(\mathbf{V}) = \left| \bigcup_{i=1}^L \mathcal{P}(V_i) \right|\,,
\end{eqnarray}
where $\mathcal{P}(M)$ and $\left| M \right|$ denote the 
power set and counting measure, respectively, of a set~$M$.
A more convenient but equivalent definition of the rank can be given in terms
of the Fourier expansion~(\ref{eqn:spin_glass_energy}), where
it is equal to the number of non-zero
coupling constants (including the $H_i$ and $F_0$).
In this section, we will first calculate
the rank for the classic neighborhoods of block, adjacent
and random type. We will then generate neighborhoods of arbitrary rank that
interpolate between these types and show that the AW-based landscape measures considered 
in the previous sections are correlated to the rank as well.

\subsubsection{Calculation of the rank}\label{sec:calc_rank}

\paragraph*{Block neighborhood:}
For the block model the rank is straightforward to obtain. Each block contains
every subset of size smaller or equal to $K$, giving a contribution
of $2^K$ to the rank. Since there are $L/K$ blocks and the empty set is counted
only once, we obtain
\begin{eqnarray}
 r_\mathrm{blc}=\frac{L}{K}\left( 2^K - 1 \right) + 1\,.
\end{eqnarray}

\paragraph*{Adjacent neighborhood:}

\begin{figure}
\centering
\includegraphics[width=0.6\textwidth]{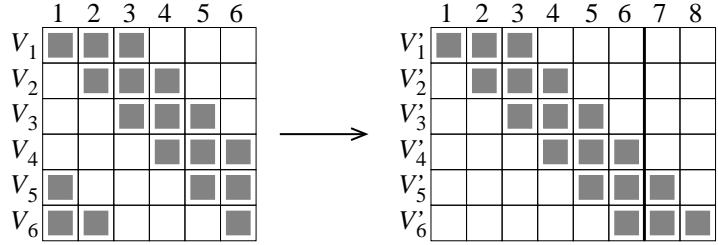}
\caption{Illustration of the calculation of the rank for the
adjacent neighborhood model. A grey square in the $i$-th row and
$j$-th column means that $V_i$
and $V_i'$, respectively, contains~$j$.
The left panel shows an adjacent neighborhood scheme
for $L=6$ and $K=3$, and the right panel shows the extended scheme for which
the rank is actually calculated.
\label{fig:adj_rank_illu}} 
\end{figure}

We will show that the rank of the adjacent neighborhood is given by 
\begin{eqnarray}
 r_\mathrm{adj} = 1 + L \cdot 2^{K-1}
\end{eqnarray}
for $K \le (L+1)/2$. For the calculation we define sets $V_i'$ which
are the same as the standard
neighborhood sets $V_i$ from (\ref{eqn:adj_nbh_set}) but without
taking the elements modulo $L$, i.e., the $V_i'$ contain elements up to 
$L+K-1$ (see figure~\ref{fig:adj_rank_illu}).
Furthermore define
\begin{eqnarray}
 \mathcal{M}' = \left\{ M \in \bigcup_{i=1}^N \mathcal{P}(V_i') \mid \min(M) \le L \right\}\,.
\end{eqnarray}
It is straightforward to count the number of elements in $\mathcal{M}'$. A set
$M \subset \mathbb{N}$ is contained in $\mathcal{M}'$ if and only if
\begin{eqnarray}
1 \le \min(M) \le L
\qquad\mathrm{and}\qquad
 \max(M) - \min(M) < K
\,.
\label{eqn:condition_subset_contained}
\end{eqnarray}
For given $a=\min(M)$ and $b=\max(M)$, there are $2^{b-a-1}$ possible sets.
Summing over all $a$ and $b$ which fulfill (\ref{eqn:condition_subset_contained}) leads to
\begin{eqnarray}
\left| \mathcal{M}' \right|
&= 1 + L + \sum_{a=1}^L \sum_{b=a+1}^{a+K-1} 2^{b-a-1}
\nonumber
\\
&= 1 + L + L \cdot \sum_{d=0}^{K-2} 2^d = 1 + L \cdot 2^{K-1}\,.
\label{eqn:derivation_adj_rank}
\end{eqnarray}
We will now show that indeed
$\left| \mathcal{M}' \right| = \left| \mathcal{M} \right| = r_\mathrm{adj}$
for $K \le (L+1)/2$. Note, however, that
$\left| \mathcal{M}' \right| > \left| \mathcal{M} \right|$ for $K > (L+1)/2$
and hence (\ref{eqn:derivation_adj_rank}) overestimates the actual rank in this case.
We define the function
$\mathrm{mod}\colon \mathcal{M}' \to \mathcal{M}$ by
\begin{eqnarray}
\mathrm{mod}(M') = \{\mod{i}{L} \mid i \in M' \}\,
\end{eqnarray}
and show that it is bijective if $K \le (L+1)/2$. In fact the $\mathrm{mod}$-function
is always surjective, but only injective if $K \le (L+1)/2$.
Let $M=\{m_1,\ldots,m_n\}~\in~\mathcal{M}$ with $m_1 < m_2 < \ldots < m_n$. Define
\begin{eqnarray}
 M' = \cases{M & $m_n-m_1 < K$\\
	\{g(m_1),\ldots,g(m_n)\} & else}
\end{eqnarray}
where
\begin{eqnarray}
 g(m) = \cases{
	m & if $m \ge K$\\
	m+L & else.
      }
\end{eqnarray}
Since $M' \in \mathcal{M}'$ and $\mathrm{mod}(M')=M$, the function
is surjective and hence $\left| \mathcal{M}' \right| \ge \left| \mathcal{M} \right|$.
Now let $A=\{a_1,\ldots,a_n\}$ and $B=\{b_1,\ldots,b_n\}$, where $a_k < a_{k+1}$ and $b_k < b_{k+1}$ for all~$k$,
be two elements of $\mathcal{M}'$ with $\mathrm{mod}(A) = \mathrm{mod}(B)$.
We will show that either $A=B$ or $K>(L+1)/2$, i.e., $\mathrm{mod}$ is injective
if $K \le (L+1)/2$. We denote by $i$ and $j$ the smallest indices that
fulfill $a_i \ge L$ and $b_j \ge L$, respectively, which means that
\begin{eqnarray}
 \mathrm{mod}(A) &= \{a_{i}-L, \ldots, a_n-L,a_1,\ldots,a_{i-1} \}\nonumber
\\
&= \{b_{j}-L, \ldots, b_n-L,b_1,\ldots,b_{j-1} \} =  \mathrm{mod}(B) \,.
\label{eqn:two_sets_proof_adj}
\end{eqnarray}
Note that the elements of $\mathrm{mod}(A)$ and $\mathrm{mod}(B)$ are written in ascending order in~(\ref{eqn:two_sets_proof_adj}).
Therefore, it is obvious that $A=B$ if $i=j$ and thus we assume without loss of generality that 
$i<j$. By comparison
of the elements one finds that $a_n-L=b_{j-i}$ and $a_1=b_{j-i+1}$. Because both $A$ and $B$ are elements
of $\mathcal{M}'$, the conditions
$a_n-a_1 = b_{j-i} - b_{j-i+1} + L < K$ and $b_{j-i+1} - b_{j-i} < K$
have to be fulfilled which finally leads to
\begin{eqnarray}
K > b_{j-i+1} - b_{j-i} > L-K
\quad\Rightarrow\quad
K > \frac{L+1}{2}\,.
\end{eqnarray}

\paragraph*{Random neighborhood:}

Due to the random choice of neighbors, the rank is a random
variable in this case, and we will calculate its expectation value.
First the probability that a given set $W$
is a subset of $V_i$ is needed.
Since $V_i$ always contains~$i$, the probability depends on whether
$i \in W$ or not. We find
\begin{equation}
 \mathbb{P}\left( W \subset V_i \mid i \in W \right) = \frac{(K-1)!}{(L-1)!} \frac{(L-m)!}{(K-m)!} =: p_m
\end{equation}
and
\begin{equation}
 \mathbb{P}\left( W \subset V_i \mid i \notin W \right) = \frac{(K-1)!}{(L-1)!} \frac{(L-1-m)!}{(K-1-m)!} = p_m \cdot \frac{K-m}{L-m}
\,,
\end{equation}
where $m$ is the number of elements of $W$.
Hence the probability $q_m$ that $W$ is contained in at least one of the
neighborhood sets is given by
\begin{eqnarray}
 q_m &= \mathbb{P}\left(\exists\,i \colon W \subset V_i\right)\nonumber\\
&= \mathbb{P}\left(\exists\,i \in W \colon W \subset V_i\right)
+ \mathbb{P}\left(\exists\,i \notin W \colon W \subset V_i\right)\nonumber\\
&= 1 - \left(1-p_m\right)^m  + \left[ 1 - \left(1-p_m \cdot \frac{K-m}{L-m}\right)^{L-m} \right]
\label{eqn:prob_set_contained}
\end{eqnarray}
for $m \geq 2$ and obviously $q_0=q_1=1$.
There are $\binom{L}{m}$ such subsets $W$ for each size $m$ and hence the
mean rank is given by
\begin{equation}
 r_\mathrm{rnd} = \sum_{m=0}^K \binom{L}{m} q_m
= 1 + L + \sum_{m=2}^K \binom{L}{m} q_m\,.
\end{equation}
The result can be simplified by using the approximation
$(1-x)^n \approx 1-nx$ in~(\ref{eqn:prob_set_contained}), which is
valid when $p_m$ is very small. This yields $q_m \approx K \cdot p_m$
and hence
\begin{eqnarray}
 r_\mathrm{rnd} &\approx 1 + L + \sum_{m=2}^K \binom{L}{m} \,  \frac{K!}{(L-1)!} \frac{(L-m)!}{(K-m)!}
 \\
&= 1 + L\,\left(2^K-K\right) = r_\mathrm{max} \,,
\end{eqnarray}
where $r_\mathrm{max}$ is the upper limit for the rank of a neighborhood with
fixed $L$ and $K$ \cite{buzas13}.

\subsubsection{Correlation between walk properties and rank.}
\begin{figure}
 \includegraphics[width=\textwidth]{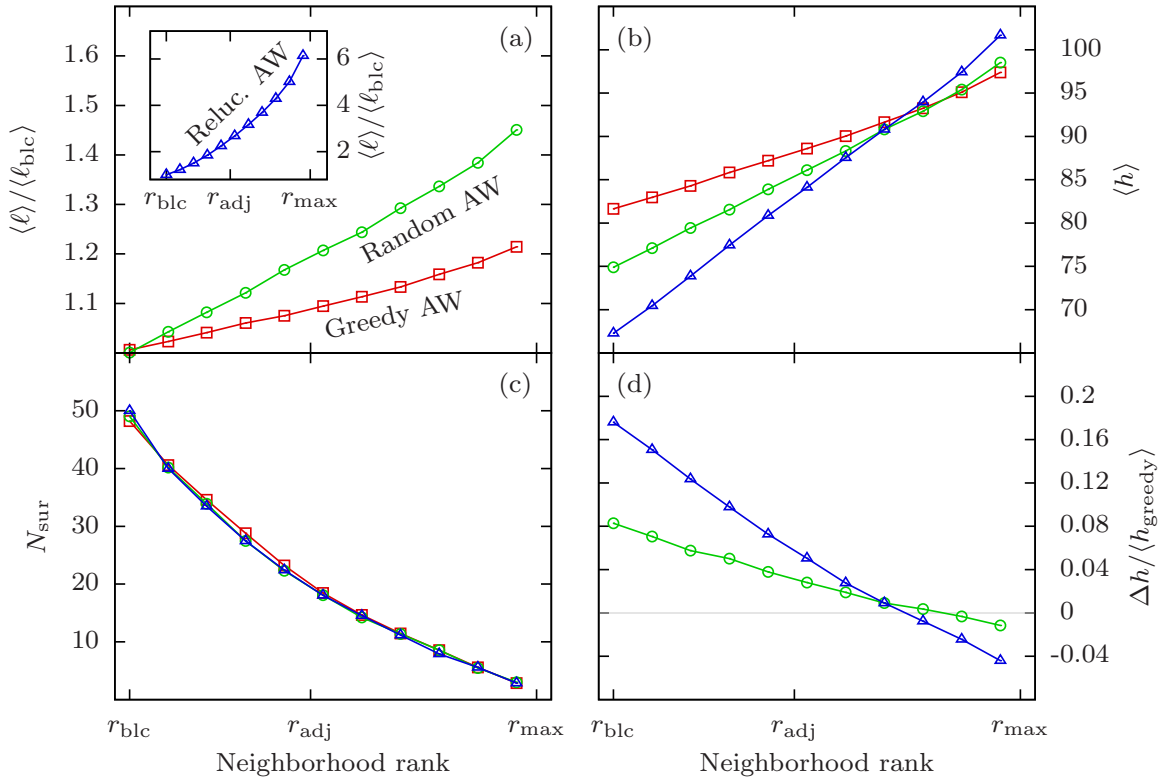}
\caption{Correlation between the rank and several properties of adaptive 
walks on an $L=128$, $K=8$ landscape. The quantities shown are (a) the
mean walk length $\langle \ell \rangle$, (b) the mean walk 
height~$\langle h \rangle$, (c) the 
number~$N_\mathrm{sur}$ of maxima at distance~$d=2$ to the
final genotype of the walk and (d) the height advantage~$\Delta h$ of
greedy adaptive walks.
Lines are for visual guidance.\label{fig:rank_correl}}
\end{figure}

Literally all quantities we analyzed in
section~\ref{sec:classic_nh} were either minimal or
maximal for block and random neighborhoods, with the values for the adjacent
model lying in between. It is therefore not surprising that
the same holds true for the rank, which is minimal
for block and maximal for random neighborhoods.
This is not a coincidence, as most quantities seem to be
generally correlated to the rank, as we are now going to show by
analyzing neighborhoods with arbitrary rank.
To generate these neighborhoods, we use the following algorithm:
\begin{enumerate}
 \item Start with a block neighborhood.\label{enum:algo_start}
 \item Choose randomly a set $V_i$, an element $n \in V_i$ with $n
   \neq i$, and replace it by another element $m \notin V_i$.\label{enum:algo_change}
 \item If the rank has been increased due to this operation,
the change in $V_i$ is accepted. Otherwise, the change is undone.
\item If no rank increasing changes are found in 1000 successive trials,
we start again at step~(\ref{enum:algo_start}).
Otherwise, we continue with step~(\ref{enum:algo_walk}).
 \item When the rank hits a prescribed threshold, an adaptive walk is performed
with the current neighborhood sets.\label{enum:algo_walk}
 \item Go to step~(\ref{enum:algo_change}).
\end{enumerate}
With this method, we can produce thousands of different neighborhood schemes with
ranks between $r_\mathrm{blc}$ and $r_\mathrm{max}$, although the
maximal rank that can be achieved in this way is usually somewhat below $r_\mathrm{max}$.
The results are shown in figure~\ref{fig:rank_correl}. 
Clearly, all quantities considered in
section~\ref{sec:classic_nh}, i.e., the
mean walk length $\langle \ell \rangle$, height~$\langle h \rangle$ and
the number~$N_\mathrm{sur}$ of local maxima at distance~$d=2$ to the
final genotype of the walk, are strongly related to the rank and
either increase or decrease monotonically with $r$. 
For both height and length it turns out that 
the different walk types react differently to variations of the rank,
whereas $N_\mathrm{sur}$, similar to the results shown in 
figure~\ref{fig:nk_clustering}(b), is largely independent of the walk type.
The length of greedy and random adaptive walks increases roughly linearly
with rank, with a larger slope for random AW's. Reluctant
walks are more susceptible to alterations of the rank, the
dependence of walk length on $r$ being stronger than linear.

Walk heights show a similar behavior as the length in terms of
the sensitivity of different walk types. Reluctant walks do show
a roughly linear dependence on the rank here, but the slope is larger
than that for random and greedy walks. Since reluctant and random walks
reach a lower height for the minimal rank $r_\mathrm{blc}$, their
height-rank curve may have a point of intersection with the curve for the
greedy walk. This point marks the threshold where reluctant and
random AW's become more successful in their ability
to find large fitness values.
Such a point exists for the landscape parameters $L=128$ and $K=8$
chosen here, but not in general. As suggested by
figure~\ref{fig:nk_all_grd-rnd}, 
an intermediate value of $K$ compared to $L$ is needed to observe this phenomenon,
and $L$ has to be sufficiently large.

\section{Summary and discussion}\label{sec:conclusion}

In this paper we studied different adaptive walk models on the
NK landscape with the focus on the differences between interaction
schemes of the NK model. In section~\ref{sec:classic_nh} we analyzed
three classic neighborhood types as well as three walk types,
resulting in nine different combinations. The picture that we obtain is
nevertheless rather simple: For each walk type,
both the mean walk length $\langle\ell\rangle$
and height $\langle h \rangle$ are largest for random,
second largest for adjacent and smallest for block neighborhoods, while
the order is reversed for the number $N_\mathrm{sur}$ of maxima surrounding
the final genotype of a walk.
Similarly, for each neighborhood type, $\langle\ell\rangle$ is
largest for reluctant, second largest for random and
smallest for greedy walks. In most situations, the opposite ordering applies
to $\langle h \rangle$, but for random neighborhoods and certain
values of $K$ and $L$ this order can be reversed and reluctant
walks become the most successful ones in terms of height.

In section~\ref{sec:rank} we showed that this picture can be extended
to more general choices of the neighborhood which can be classified in
terms of the rank. Block, adjacent and random neighborhoods are just
examples of schemes with low, medium and high rank, respectively.
Our findings concerning the relation between walk length and rank are
consistent with results from previous work, since $\ell$ would be
expected to be related to the density of local maxima which decreases
slightly with increasing rank \cite{buzas13}. In this sense,
an increasing rank decreases the ruggedness of a landscape.
Note that this is also consistent
with another measure of ruggedness, namely the probability to find
an accessible path to the global maximum, which 
was found to be largest for random,
second largest for adjacent and smallest for block neighborhoods
\cite{schmiegelt13}.

If the number of local maxima decreases with increasing rank, one
would expect the same trend for the number of maxima $N_\mathrm{sur}$
surrounding a given maximum. Though this is true, the effect on $N_\mathrm{sur}$ is much
stronger than that on the walk length and the number of maxima, which indicates that the
rank affects the distribution of maxima in the landscape more
substantially than their density. Maxima become
much more isolated with increasing rank. The fact that $N_\mathrm{sur}$
hardly depends on the walk type suggests that this is true
for typical local maxima and not only for those found by adaptive walks.

The rank thus appears to be a powerful tool for the characterization and description
of neighborhood schemes, 
but so far it lacks explanatory power.
In fact, it is quite surprising that the ruggedness decreases with the rank
for fixed $L$ and $K$, since the opposite is true if the neighborhood
type is fixed and the rank is increased due to an increase of $K$ \cite{manukyan14}.
Because of the difficulty of analytical approaches to the NK model for
neighborhoods that are not block-like as well as the impossibility to
exhaustively enumerate the entire landscape for large $L$, we are for now restricted
to indirect measurements using adaptive walks.
Nevertheless, we showed that the model is rich in interesting and
non-trivial phenomena and hope that the dependence of landscape properties
on interaction schemes will be investigated more frequently in future work.

With regard to the application of probabilistic models for the interpretation
of empirical fitness landscapes, our work highlights the importance of
developing refined measures of genetic interactions that go beyond the
summary statistics of fitness landscape ruggedness considered in most
previous studies \cite{szendro13,devisser14}. Moreover, our
demonstration that different types of adaptive walks respond differently 
to the structure of these interactions suggests
a new methodology for exploring high-dimensional empirical data sets,
where adaptive walks have so far been employed only for estimating the
correlation length and overall density of local maxima in the
landscape \cite{kouyos12}.

\ack
We thank Cristian Giardina for pointing us to the literature on reluctant adaptive 
walks. This work was supported by DFG within SPP 1590
\textit{Probabilistic structures in evolution}.


\appendix
\setcounter{section}{0}

\section{Adaptive walks on the HoC landscape}\label{apx:height_of_walks}

Here we derive several quantities for adaptive walks
on the uncorrelated House-of-Cards landscape. Although the properties
of the walks do not depend on the underlying fitness distribution, for
convenience the fitness values are assumed to be uniformly distributed
on the interval $[0,1]$.

\subsection{Height of greedy adaptive walks}\label{apx:hoc_greedy}

To calculate the mean walk height for greedy adaptive walks
is rather simple. If the walk has a length $\ell$, the population sees
in total $(\ell+1)\cdot L$ genotypes and chooses the one with largest
fitness. The mean value of the largest of $n$ i.i.d. uniform
random variables is given by $M_n = n/(n+1)$, the probability that the
walk has length $\ell$ is given by $P_\ell = \ell/(\ell+1)!$ \cite{orr03} and
hence the mean walk height is given by
\begin{eqnarray}
\langle h \rangle &= \sum_{\ell=0}^\infty M_{(\ell+1) \cdot L} \cdot P_\ell
    = 1-\sum_{\ell=0}^\infty \frac{1}{(\ell+1)\cdot L + 1} \cdot \frac{\ell}{(\ell+1)!}
\\
&\approx 1-\sum_{\ell=0}^\infty \frac{1}{(\ell+1)\cdot L} \cdot \frac{\ell}{(\ell+1)!}
= 1-\frac{\alpha_\mathrm{grd}}{L}
\label{eqn:hoc_approx_grd}
\end{eqnarray}
where
\begin{eqnarray}
 \alpha_\mathrm{grd} = \sum_{\ell=0}^\infty \frac{\ell}{(\ell+1)\cdot (\ell+1)!} = 0.4003\ldots
\end{eqnarray}

\subsection{Height of random adaptive walks}\label{apx:hoc_random}

The probability density $q_L(x)$ for the height of random adaptive walks on a 
uniformly distributed HoC landscape is
known \cite{flyvbjerg92}, so we will just compute its average.
The density function of the height is given by
\begin{eqnarray}
 q_L(x)=x^{L-1} \cdot \exp\left( \sum_{k=1}^{L-1} \frac{x^k}{k} \right)\,.
\end{eqnarray}
In order to compute the mean value, we define
\begin{eqnarray}
 r_L(x):=\frac{q_L(1-x/L)}{q_L(1)} = \left( 1 - \frac{x}{L} \right)^{L-1} \exp \left( \sum_{k=1}^{L-1} \frac{\left(1-\frac{x}{L}\right)^k - 1}{k} \right)\,.
\end{eqnarray}
The exponent can be written as
\begin{eqnarray}
\xi(x, L) &:=\sum_{k=1}^{L-1} \frac{\left(1-\frac{x}{L}\right)^k - 1}{k}
\\
&= \sum_{k=1}^{L-1} \sum_{s=1}^k \frac{(k-1)!}{s!\cdot(k-s)!} (-1)^s \left( \frac{x}{L} \right)^s
\\
&= \sum_{s=1}^{L-1} \sum_{k=s}^{L-1} \frac{(k-1)!}{s!\cdot(k-s)!} (-1)^s \left( \frac{x}{L} \right)^s
\\
&= \sum_{s=1}^{L-1} \frac{(-1)^s\,x^s}{s \cdot s!} \cdot \frac{(L-1)!}{L^s\,(L-s-1)!}
\end{eqnarray}
The second factor is smaller than 1 and bounded from below by $1-s^2/L$, i.e.,
\begin{eqnarray}
 \sum_{s=1}^{L-1} \frac{(-1)^s\,x^s}{s \cdot s!}
 \ge \xi(x, L) 
 \ge \left( \sum_{s=1}^{L-1} \frac{(-1)^s\,x^s}{s \cdot s!} \right) - R(x,L)
\end{eqnarray}
with the remainder term
\begin{eqnarray}
 R(x,L)
= -\frac{1}{L} \sum_{s=1}^{L-1}  \frac{(-1)^s\,x^s}{(s-1)!}.
\end{eqnarray}
Since
\begin{eqnarray}
\lim_{L \to \infty} L R(x,L)   
= - \sum_{s=1}^{\infty} \frac{(-1)^s\,x^s}{(s-1)!}
= x\,e^{-x}
\end{eqnarray}
is finite, $R(x,L)$ tends to zero for $L \to \infty$. By the squeeze theorem it follows that
\begin{eqnarray}
\lim_{L \to \infty} \xi(x, L) 
= \sum_{s=1}^{\infty} \frac{(-1)^s x^s}{s \cdot s!}
= -\left( \log (x)+\Gamma(0,x)+\gamma \right)
\end{eqnarray}
where $\gamma\approx0.5772...$ is the Euler-Mascheroni constant and 
$\Gamma (a,z)=\int _z^{\infty } t^{a-1} e^{-t}\,\mathrm{d}t$ is the incomplete
gamma function. Hence the function series $r_L$ converges for
$L \to \infty$ to a non-degenerate limiting function
\begin{eqnarray}
 r_\infty(x) =& \exp\left( -x-\log(x)-\Gamma(0,x)-\gamma) \right)\,.
\end{eqnarray}
Since $r_\infty$ does not depend on $L$, we can extract the $L$-dependence of $q$.
With the substitution $x=1-y/L$ we find
\begin{eqnarray}
\langle h \rangle &= \int_0^1 x\,q_L(x)\,\mathrm{d}x 
= \frac{1}{L} \int_0^L \left(1-\frac{y}{L} \right) \cdot q_L\left(1-\frac{y}{L} \right)\,\mathrm{d}y
\\
&= 1 - \frac{q_L(1)}{L^2} \int_0^L y\,r_L(y) \,\mathrm{d}y
\approx 1 - \frac{e^\gamma}{L} \int_0^\infty y\,r_\infty(y) \,\mathrm{d}y
\\
&= 1 - \frac{\alpha_\mathrm{rnd}}{L}
\label{eqn:hoc_approx_rnd}
\end{eqnarray}
with
\begin{eqnarray}
 \alpha_\mathrm{rnd} = \int_0^\infty \, \exp\left( -x-\Gamma(0,x) \right) \,\mathrm{d}x = 0.6243\ldots
\end{eqnarray}
This result was previously obtained in \cite{macken91}.

\subsection{Reluctant walks}\label{apx:hoc_reluctant}
\begin{figure}[b!]
 \includegraphics[width=\textwidth]{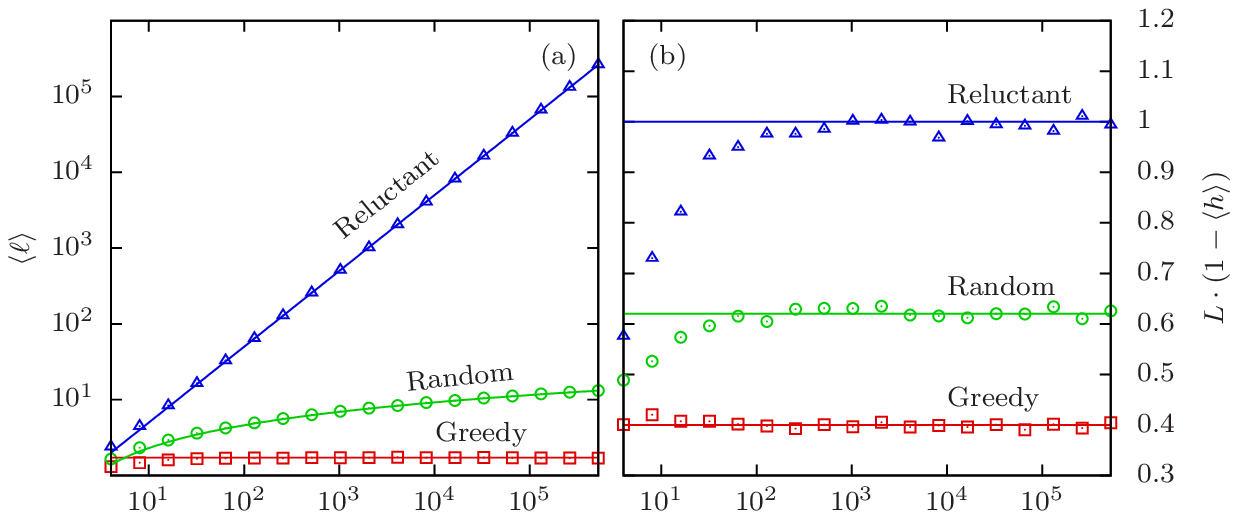}
\caption{(a) Mean walk length $\langle \ell \rangle$ and (b) height $\langle h \rangle$
on the House-of-Cards landscape. The numerical results suggest  that 
$\langle \ell \rangle=\frac{L}{2}$ and $\langle h
\rangle=1-\frac{1}{L}$ for the reluctant walk.
\label{fig:hoc_length_height}}
\end{figure}

For the reluctant walk on the HoC landscape, we find numerically that the length is asymptotically
given by $\langle\ell\rangle=L/2$ and the height by $\langle h\rangle=1-1/L$
(see figure~\ref{fig:hoc_length_height}). These results are plausible
within the Gillespie approximation \cite{gillespie83,orr02}, 
a simplified setting where the entire adaptive walk proceeds among a
single set of $L$ random fitness values; in other words, the creation
of a new neighborhood of independently drawn fitness values after each
step is neglected. Somewhat surprisingly, the Gillespie approximation has been shown to
correctly reproduce the leading order $\log L$-behavior for the length
of random and natural AW's \cite{orr02,neidhart11}. For the greedy AW it trivially
predicts $\ell = 1$, which is also rather close to the exact result
$\langle\ell\rangle = e-1 \approx 1.7183$. Within the Gillespie approximation, a
reluctant walk visits all sites of the neighborhood in order of
increasing fitness, and the walk length is equal to the rank of the initial fitness
among the other $L$ fitness values in the neighborhood minus one. It
follows that the length is $L/2$ on average.

As a starting point for a systematic treatment of reluctant AW's on
the HoC landscape, we derive a recurrence relation for the quantity
\begin{eqnarray}
 P_{\ell}(x) := \mathbb{P}(\text{fitness in $[x, x+dx]$ after $\ell$ steps})\,,
\end{eqnarray}
following the procedure of Flyvbjerg and Lautrup \cite{flyvbjerg92}.
The recurrence relation in general reads
\begin{eqnarray}
P_{\ell+1}(x) = \int_{-\infty}^{x}
P_{\ell}(y)\,\gamma(y \to x)
\,\mathrm{d}y
\label{eqn:general_recurrence}
\end{eqnarray}
where $\gamma(y \to x)$ is the probability density of the smallest of $L$ random
variables that is larger than $y$. For uniform random variables and
conditioned on there being $k > 0$ random variables
larger than $y$, the density is given by
\begin{eqnarray}
 \gamma(y \to x \mid k) = \frac{k}{1-y}\,\left(1 - \frac{x-y}{1-y} \right)^{k-1}
\end{eqnarray}
and hence
\begin{eqnarray}
\gamma(y \to x) 
&= \sum_{k=1}^{L} \binom{L}{k} \, (1-y)^k \,y^{L-k} \, \gamma(y \to x \mid k)
\\
&= \sum_{k=1}^{L} \binom{L}{k}\,k\,(1-x)^{k-1}\,y^{L-k}
\\
&= L \,(1-x+y)^{L-1}
\,.
\end{eqnarray}
Then equation (\ref{eqn:general_recurrence}) becomes
\begin{eqnarray}
 P_{\ell+1}(x) = L \int_{0}^{x} P_{\ell}(y) \, (1 - x + y)^{L-1}\,\mathrm{d}y
\label{Eq:relrecur}
\end{eqnarray}
from which quantities of interest like walk lengths and heights could
in principle be extracted. However, so far we have not succeeded in
solving the recursion (\ref{Eq:relrecur}). 

\subsection{Walk height for fitness distributions in the Gumbel class}

So far, the calculations of the mean value of the walk height $h$ were based
on the assumption that the fitness values are uniformly distributed.
Obviously, fitness values drawn from another continuous distribution
can always be transformed to the uniform
case (and vice versa), since for a random variable $X$ with cumulative
distribution function~$Q$ the distribution of $Q(X)$ is
uniform.
Therefore, the transformed height $Q(h)$ for an arbitrary
distribution has the same probability
density function $q(x)$ as the height in the uniform case.
One could in principle get to the mean height $h$ by
\begin{eqnarray}
 \langle h \rangle = \langle Q^{-1}(Q(h)) \rangle
 = \int_0^1 Q^{-1}(x)\,q(x)\,\mathrm{d}x \,,
\end{eqnarray}
but in practice this integral can be cumbersome to evaluate. However,
for fitness values drawn from a distribution in the Gumbel class of
extreme value theory, e.g.,
a Gaussian distribution, there is a simple approximation for the
relation between $h$ and $Q(h)$.

\begin{figure}
\centering
 \includegraphics[width=0.8\textwidth]{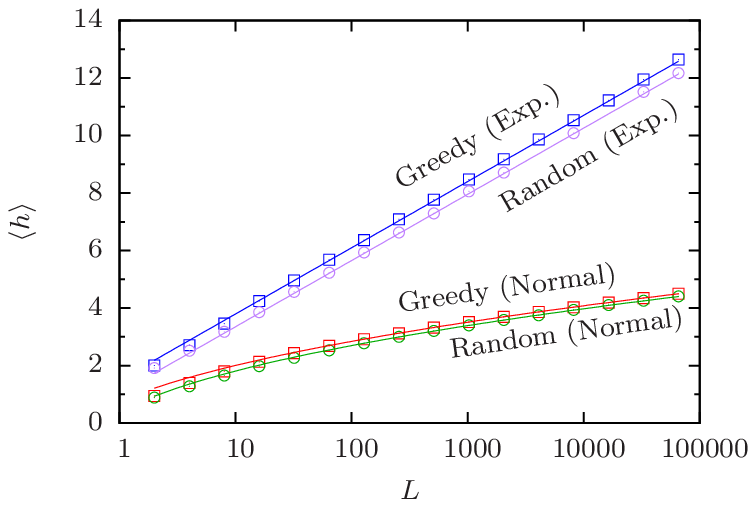}
\caption{Height of random and greedy adaptive walks on the HoC
  landscape with fitness values distributed according to a standard normal
distribution and a standard exponential distribution, respectively.
Symbols correspond to simulation results, lines correspond to~(\ref{eqn:hoc_approx_gauss}).
\label{fig:hoc_approx_nonuni}}
\end{figure}

Let $X_1,\ldots,X_n$ be i.i.d. random variables with cumulative distribution
function $Q=1-\exp(-\lambda\,x)$ and $M_n = \max(X_1,\ldots,X_n)$.
The mean value of $Q(M_n)$ is given by
\begin{eqnarray}
 \langle Q(M_n) \rangle = 1 - \frac{1}{n+1}
\end{eqnarray}
whereas the mean value of $M_n$ is given by \cite{dehaan06}
\begin{eqnarray}
 \lambda \langle M_n \rangle = H_n = \sum_{k=1}^n \frac{1}{k} = \log n + \gamma + \mathcal{O}\left(\frac{1}{n}\right)
\end{eqnarray}
where $H_n$ is the $n$-th harmonic number. This yields
\begin{eqnarray}
 Q( \langle M_n \rangle ) \approx 1 - \frac{e^{-\gamma}}{n} \approx 1 - e^{-\gamma} \left( 1 - \langle Q(M_n) \rangle \right)\,.
\label{eqn:max_of_rnd_var}
\end{eqnarray}
Because the Pickands-Balkema-de Haan theorem \cite{dehaan06,balkema74,pickands75} states that the tail of a distribution from
the Gumbel class is well described by the exponential distribution, this approximation
is also valid for more general choices of $Q$ if $n$ is large.
We now assume that $h$ behaves statistically like the maximum of $n$ random
variables, i.e., like $M_n$. Although $n$ is not necessarily known in the case
of adaptive walks, we can still use~(\ref{eqn:max_of_rnd_var}) to obtain
\begin{eqnarray}
\langle h \rangle \approx  Q^{-1} \left[ 1 - e^{-\gamma} \left( 1 - \langle Q(h) \rangle \right)  \right]
= Q^{-1} \left[ 1 - \frac{\alpha \, e^{-\gamma}}{L} \right]\,,
\label{eqn:hoc_approx_gauss}
\end{eqnarray} 
where $\alpha$ is the factor depending on the walk type that was
derived above. Despite the somewhat uncontrolled nature of the approximation, the result is
quite precise as shown in figure~\ref{fig:hoc_approx_nonuni}.


\section{Number of maxima in the adjacent NK-model with $K=2$}\label{sec:nummaximak1}
In the adjacent NK model with $K=2$, the fitness $F(\sigma)$ of a genotype
$\sigma \in \mathbb{H}_L$ is given by
\begin{eqnarray}
 F(\sigma)=\sum_{i=1}^L \eta_i (\sigma_{i}, \sigma_{i+1})\,,
\end{eqnarray}
where $\sigma_{L+1}=\sigma_{1}$ and the $\eta_i (\sigma, \tau)$ are random
numbers independently drawn from a distribution with density function $f$
for each $i$, $\sigma$ and $\tau$, i.e., one needs $4L$ random numbers to
specify the whole model. In the following we will show that, if the random
numbers are drawn from a Gamma distribution with shape parameter $p=1/2$, 
the mean number $N_\mathrm{max}$ of local maxima in such a landscape is given by
\begin{eqnarray}
 N_\mathrm{max} = (2\,\lambda_{+})^L + (2\,\lambda_{-})^L \,,
\end{eqnarray}
where
\begin{eqnarray}
 \lambda_\pm = \frac{1}{6} \left[ 3 - \sqrt{3} \pm \sqrt{6 \, \left( \sqrt{3}-1 \right) } \right] \,.
\label{eqn:eigenvalues}
\end{eqnarray}
In order to derive this result, we consider a specific genotype $\sigma$
which without loss of generality can be chosen as the all-zero genotype $\sigma = (0,\ldots,0)$,
and calculate the probability $P_\mathrm{max}$ that it is a local optimum.
Its fitness is determined by the sum of $L$ random numbers, which will be 
denoted by $x_i=\eta_i(0,0)$ and is fixed in the following. If a mutation
occurs at position $j$ of the genome, the contributions $x_{j-1}$ and $x_j$ (with $x_0=x_L$)
are replaced by two new random variables $x'_{j-1}$ and $x'_{j}$, respectively.
Obviously, if $\sigma$ is a local optimum, $x'_{j-1} + x'_{j} < x_{j-1} + x_{j}$ must
hold true. Since the $x_i$ are fixed, this probability can be written as
\begin{eqnarray}
\mathbb{P}(x'_{j-1} + x'_{j} < x_{j-1} + x_{j}) = \tilde{F}(x_{j-1} + x_{j}) \,,
\end{eqnarray}
where
\begin{eqnarray}
 \tilde{F}(x) = \int_{-\infty}^{x} \left( \int_{-\infty}^{\infty} f(z)\,f(y-z) \,\mathrm{d}z \right) \,\mathrm{d}y
\end{eqnarray}
is the cumulative distribution function of the convolution of two random variables
drawn from $f$. Note that the $x'_i$ which can occur due to mutations are independent
and therefore the probability that $\sigma$ is a maximum [for fixed
$\underline{x} = (x_1, \ldots, x_L)$] is given by
\begin{eqnarray}
 P_\mathrm{max}(\underline{x}) = \tilde{F}(x_1 + x_2) \, \tilde{F}(x_2 + x_3)
\cdots \tilde{F}(x_{L-1} + x_{L}) \, \tilde{F}(x_{L} + x_{1})\,.
\end{eqnarray}
The actual probability $P_\mathrm{max}$ can then be obtained by integrating over all values
of $\underline{x}$, i.e.,
\begin{eqnarray}
 P_\mathrm{max} = \int_{\mathbb{R}^L} \prod_{n=1}^L  \left(
f\left(x_n \right)\,\tilde{F}\left(x_n + x_{n+1}\right)
\right)
\,\mathrm{d}^L\underline{x}
\,.\label{eqn:general_probability}
\end{eqnarray}
This integral can be solved exactly if the contributions are drawn from
a Gamma distribution with shape parameter $p=1/2$, i.e., the density
function $f$ is given by
\begin{eqnarray}
 f(x) = \frac{\exp(-x)}{\sqrt{\pi\,x}}
\end{eqnarray}
for $x > 0$ and zero otherwise. 
The sum of two random variables drawn from this distribution is
exponentially distributed, i.e.,
\begin{eqnarray}
 \tilde{F}(x)= 1 - e^{-x}
\end{eqnarray}
and equation~(\ref{eqn:general_probability}) becomes
\begin{eqnarray}
 P_\mathrm{max} = \frac{1}{\sqrt{\pi^L}}
\int_{\mathbb{R}_{+}^L}
\prod_{n=1}^{L}
\left(
\frac{\exp(-x_n)}{\sqrt{x_n}} \,
\left( 1 - e^{-x_n - x_{n+1}} \right)
\right)
\mathrm{d}^L\underline{x}
\,.\label{eqn:probability_for_gamma}
\end{eqnarray}
Expanding the product yields
\begin{eqnarray}
&\prod_{n=1}^{L} \left( 1 - e^{-x_n - x_{n+1}} \right)
= \sum_{\sigma} \prod_{n=1}^{L}
(-1)^{\sigma_n} e^{-x_n (\sigma_{n-1} + \sigma_{n})} 
\,,\label{eqn:expand_stuff}
\end{eqnarray}
where the sum goes over all $\sigma \in \{0,1\}^L$.
Inserting (\ref{eqn:expand_stuff}) into (\ref{eqn:probability_for_gamma}) gives
\begin{eqnarray}
P_\mathrm{max} &=
\frac{1}{\sqrt{\pi^L}} 
\int_{\mathbb{R}_{+}^L}
\sum_{\sigma} \prod_{n=1}^{L}
\frac{(-1)^{\sigma_n}}{\sqrt{x_n}}
e^{-x_n (\sigma_{n-1} + \sigma_{n} + 1)} 
\,\mathrm{d}^L\underline{x}
\\
&=\frac{1}{\sqrt{\pi^L}}  \sum_{\sigma} \prod_{n=1}^{L} (-1)^{\sigma_n} 
\int_{0}^{\infty} \frac{\exp({-x (\sigma_{n-1} + \sigma_{n} + 1)})}{\sqrt{x}}\,\mathrm{d}x
\\
&= \sum_{\sigma} \prod_{n=1}^{L}
\frac{(-1)^{\sigma_n}}{\sqrt{\sigma_{n-1}+\sigma_{n}+1}}
=
\sum_{\sigma} \prod_{n=1}^{L} T_{\sigma_{n}, \sigma_{n+1}}
\label{eqn:looks_like_partition}
\end{eqnarray}
with the matrix
\begin{eqnarray}
T =
\left(\begin{array}{cc}
1 & -\frac{1}{\sqrt{2}} \\
\frac{1}{\sqrt{2}} & -\frac{1}{\sqrt{3}}
\end{array}\right)
\,.
\end{eqnarray}
Equation~(\ref{eqn:looks_like_partition}) is of the form of a partition function
of a spin chain and $T$ is the corresponding transfer matrix. Using this
analogue, one can write
\begin{eqnarray}
P_\mathrm{max} = \mathrm{Tr}\left( T^L \right) = \lambda_{+}^L + \lambda_{-}^L
\end{eqnarray}
where $\lambda_{\pm}$ are the eigenvalues of $T$ which are given by equation~(\ref{eqn:eigenvalues}).
The final result is obtained by multiplying the probability $P_\mathrm{max}$ with the
total number $2^L$ of genotypes which yields
\begin{eqnarray}
  N_\mathrm{max} = P_\mathrm{max} \cdot 2^L = (2 \lambda_{+})^L + (2 \lambda_{-})^L
\,.
\end{eqnarray}
For large $L$ the behavior is governed by the larger eigenvalue $\lambda_{+} = 0.5606\ldots$.

\section*{References}
\bibliographystyle{iopart-num}
\bibliography{lit}

\end{document}